\documentclass{bmcart}

\usepackage{amsthm,amsmath}
\usepackage[utf8]{inputenc} 

\usepackage{graphicx}

\usepackage{endnotes}
\let\footnote=\endnote

\startlocaldefs

\usepackage{hyperref}
\usepackage{color}
\usepackage{url}
\usepackage[caption=false,font=footnotesize]{subfig}

\def\aplt{\ {\raise-.5ex\hbox{$\buildrel<\over\sim$}}\ }

\endlocaldefs

\begin{document}

\begin{frontmatter}

\begin{fmbox}
\dochead{Research}


\title{Sapporo2: A versatile direct $N$-body library}


\author[
   addressref={aff1},                   
   corref={aff1},                       
   email={bedorf@strw.leidenuniv.nl}   
]{\inits{JB}\fnm{Jeroen} \snm{B\'edorf}}
\author[
   addressref={aff2},
]{\inits{EG}\fnm{Evghenii} \snm{Gaburov}}
\author[
   addressref={aff1},                   
   email={spz@strw.leidenuniv.nl}   
]{\inits{SPZ}\fnm{Simon} \snm{Portegies Zwart}}


\address[id=aff1]{
  \orgname{Leiden Observatory, Leiden University}, 
  \street{P.O. Box 9513 },                     %
  \postcode{NL-2300 RA}                         
  \city{Leiden},                              
  \cny{the Netherlands}                                    
}
\address[id=aff2]{%
  \orgname{SurfSARA},
  \street{P.O. Box 94613},
  \postcode{1090 GP}
  \city{Amsterdam},
  \cny{the Netherlands}
}



\end{fmbox}


\begin{abstractbox}

\begin{abstract} 


Astrophysical direct $N$-body methods have been one of the first 
production algorithms
to be implemented using NVIDIA's {\tt CUDA} architecture. Now, 
almost seven years later, the GPU is the most used accelerator device
in astronomy for simulating stellar systems. In this paper we
present the implementation of the {\tt Sapporo2} $N$-body library,
which allows researchers to use the GPU for $N$-body simulations
with little to no effort. The first version, released five years ago,
is actively used, but lacks advanced features and versatility
in numerical precision and support for higher order integrators.
In this updated version we have rebuilt the code
from scratch and added support for {\tt OpenCL}, multi-precision 
and higher order integrators. We show how to tune these 
codes for different GPU architectures and present 
how to continue utilizing the GPU optimal even
when only a small number of particles ($N < 100$) is integrated.
This careful tuning allows {\tt Sapporo2} to be faster than 
{\tt Sapporo1} even with the added options and double
precision data loads. The code runs on a range 
of NVIDIA and AMD GPUs in single and double precision accuracy.
With the addition of {\tt OpenCL} support the library is also able 
to run on CPUs and other accelerators that support {\tt OpenCL}. 

\end{abstract}




\end{abstractbox}
%

\end{frontmatter}



\section{Background}

The class of algorithms, commonly referred to as direct $N$-body algorithms is still 
one of the most commonly used methods for simulations in astrophysics. These algorithms are relatively 
simple in concept, but can be applied to a wide range of problems. From the simulation 
of few body problems, such as planetary stability to star-clusters and even small
scale galaxy simulations. 
However, these algorithms are also computationally expensive as they scale 
as $O(N^2)$. This makes the method unsuitable for large $N$ ($>10^6$), for these
large $N$ simulations one usually resorts to a lower precision method like the Barnes-Hut 
tree-code method~\cite{1986Natur.324..446B} or the Particle Mesh method that both scale
as $O(N \log N)$ (e.g.~\cite{1969JCoPh...4..306H, 1981csup.book.....H}). 
These methods, although faster, are also notably less accurate and not suitable for 
simulations that rely on the high accuracy that direct summation, coupled with higher order integrators, offer. 
On the other end of the spectrum you can find even higher accuracy methods 
which uses arbitrary precision~\cite{2014ApJ...785L...3P}. The work
of~\cite{2014ApJ...785L...3P} indicates that the accuracy offered by
the default (double precision) 
direct $N$-body methods is sufficient for most scientific problems.

The direct $N$-body algorithm is deceivingly simple, in the fundamental form it performs $N^2$ gravitational
computations, which is a parallel problem that can be efficiently implemented
on almost any computer architecture with a limited amount of code lines. 
A number of good examples can be found on the {\tt Nbabel.org} website. 
This site contains examples of a simple $N$-body simulation code 
implemented in a wide range of programming languages.
However, in practice there are many variations of the algorithms in use, 
with up to eighth order integrations~\cite{2008NewA...13..498N}, algorithmic extensions such as  
block time-stepping~\cite{1986LNP...267..156M}, 
neighbour-schemes~\cite{1973JCoPh..12..389A}, 
see~\cite{2012EPJST.210..201B} and references therein
for more examples. These variations transform the simple
$O(N^2)$ shared time-step implementation in a complex method,
where the amount of parallelism can differ per time-step. 
Especially the dynamic block time-stepping method adds complexity
to the algorithm, since the number of particles that participate
in the computations changes with each integration step. 
This variable number of particles involved in computing forces 
requires different parallelisation strategies.
In the worst case, there is only
one particle integrated, which eliminates most of the standard parallelisation
methods for $N^2$ algorithms. 
There is extensive literature on high performance direct $N$-body methods with the 
first being described in 1963~\cite{1963MNRAS.126..223A}. The method has been efficiently
implemented on parallel machines~\cite{1986LNP...267..156M}, vector machines~\cite{1988CeMec..45...77H}
and dedicated hardware such as the GRAPE's~\cite{1998sssp.book.....M}. 
For an overview we refer the interested reader to the following 
reviews~\cite{2012EPJST.210..201B, 2003gmbp.book.....H, 2011EPJP..126...55D}.
Furthermore, there has been extensive work on accelerating $N$-body methods
using GPUs. There have been several $N$-body libraries to ease the development 
of $N$-body integrators that use the GPU. The first library that offered 
support for the GRAPE API 
was {\tt Kirin}~\cite{2008NewA...13..103B}, however this library only 
supports single precision and is therefore less accurate than the GRAPE. 
With the introduction of the {\tt Yebisu} library~\cite{keigo_thesis} there
was support for double-single precision\footnote{In this precision, the number of significant 
digits is 14 compared to 16 in IEEE double precision. Using a pair of floating point numbers double precision 
accuracy is approximated through single precision floating point operations},
which achieved accuracy comparable
to the GRAPE. The library also featured support for fourth and sixth order
Hermite integrators in combination with minimized data send by performing
the prediction on the GPU. This library, however, is not compatible with 
the GRAPE API and only supports a single GPU.  In our previous work
{\tt Sapporo1}~\cite{Gaburov2009630}, we added support for multiple GPUs
in combination with the GRAPE API and double-single precision. 
Apart from libraries there are also $N$-body integrators that 
come with built-in support for GPU hardware. For example in~\cite{2011hpc..conf....8B},
the authors combine {\tt Yebisu} and {\tt phiGRAPE~\cite{2007NewA...12..357H}} in the 
new {\tt phiGPU} code. This code is able to run on multiple GPUs and supports
up to eighth order accuracy. In~\cite{2013JCoPh.236..580C, 2013CoPhC.184.2528C},
the authors introduce the {\tt HiGPUs} $N$-body code. This standalone 
code contains a sixth order integrator, and supports CUDA, OpenCL
and IEEE-754 double precision accuracy.  Finally, there is {\tt NBODY6} which
uses GPU acceleration together with an Ahmad-Cohen neighbour 
scheme~\cite{1973JCoPh..12..389A, 2012MNRAS.424..545N}.

In this paper we present our direct $N$-body library, {\tt Sapporo2},
since we focus on the library we will not make a full comparison with the standalone 
software packages mentioned above.
The library contains built-in support for the second order leap-frog (GRAPE-5), 
fourth order Hermite (GRAPE-6) and sixth order Hermite integrators.
The numerical precision can be specified at run time and depends on
requirements for performance and accuracy. Furthermore, the library can keep track of the nearest 
neighbours by returning a list containing all particles within a certain radius.
Depending on the available hardware the library operates with CUDA and OpenCL, and has the
option to run on multiple-GPUs, if installed in the same compute node. 
The library computes the gravitational force on particles
that are integrated with block time-step algorithms.
However, the library can trivially be applied to any other $O(N^2)$ 
particle method by replacing the force equations. 
For example, methods that compute the Coulomb 
interactions~\cite{VanGorp2011192} or molecular 
dynamics~\cite{12392506} use  similar methods as presented in this work.

\section{Methods}
\label{Sapporo2:Sect:Method}

With Graphic Processing Units (GPUs) being readily available in the computational astrophysics community
for  over 5 years we will defer a full description of their specifics and 
peculiarities~\cite{2012EPJST.210..201B, 2008NewA...13..103B,Nyland_nbody,CUDAGuide5.5}.
Here we only give a short overview to stage the context for the following sections.
In GPU enabled programs we distinguish two parts of code. 
The `host' code, used to control the GPU, is executed on the CPU; whereas the `device' 
code, performing the majority of the computations, is executed on the GPU.
Each GPU consists of a set of multiprocessors 
and each of these multiprocessors contains a set of computational units. We send 
work to the GPU in blocks for further processing
by the multiprocessors. In general a GPU requires a large amount of these blocks to
saturate the device in order to hide most of the latencies
that originate from communication with the off-chip memory. 
These blocks contain a number of threads that perform computations.
These threads are grouped together in `warps' for NVIDIA machines or `wavefronts' on AMD
machines. Threads that are grouped together 
share the same execution path and program counter. The smaller the number of threads 
that are grouped the smaller the impact of thread divergence. On current devices
a warp consists of 32 threads and a wavefront contains 64 threads. This difference 
in size has effects on the performance (see Section~\ref{Sapporo2:Sect:Results}).

\subsection{Parallelisation method}
\label{Sapporo2:Sect:Method:Par}

To solve the mutual forces for an $N$-body system the forces exerted by 
the $j$-particles (sources) onto the $i$-particles (sinks) have to be 
computed. Depending on the used algorithm the sources and sinks can 
either belong to the same or a completely different particle set. 
Neither is it required that these sets have the same dimensions. 
In worst case situations this algorithm scales as $O(N^2)$, but since
each sink particle can be computed independently it is trivial to parallelise 
within a single time-step. 
The amount of parallelism, however, depends on the number of sink
particles. For example, in high precision gravitational direct 
$N$-body algorithms that employ block time-stepping the number
of sink particles ranges between 1 and $N$. 
In general the number of sinks is smaller than the number of sources, 
because only the particles of which the position and velocity 
require an update are integrated~\cite{1986LNP...267..156M}.
As a consequence the amount of available parallelism in this algorithm
is very diverse, and depends directly on the number of active sink particles.

Currently there are two commonly used methods for solving $N^2$ like 
algorithms using GPUs.  
The first performs parallelisation over the sink particles 
~\cite{2007astro.ph..3100H, 2008NewA...13..103B, Nyland_nbody}  
which launches a separate compute thread for each sink particle. 
This is efficient when 
the number of sinks is large ($> 10^4$), because then the number of compute
threads is sufficiently high to saturate the GPU. However, when the number of sink particles 
is small ($\leq 10^4$) there are not enough active compute threads to 
hide the memory and instruction latencies. As a result, the GPU will be under utilized
and only reaches a fraction of the available peak performance. 
We expect that future devices require an even larger number of running threads
to reach peak performance, in which case the number of sink particles has to be even larger to 
continuously saturate the device.
However, adjusting the number of sink particles to keep
parallel efficiency is not ideal, because then one artificially 
increases the amount of work (and run time) in favor of efficiency. 
Therefore, a second 
method was introduced in {\tt Sapporo1}~\cite{Gaburov2009630} which takes 
a slightly different approach. In {\tt Sapporo1} we parallelize over 
the source particles and keep the number of sink particles that is concurrently integrated
fixed to a certain number.
The source particles are split into subsets, each of which
forms the input against which a set of sink particles is integrated.
The smaller the number of sink particles the more subsets of source particles
we can make.
It is possible to saturate the GPU with enough subsets, 
so if the product of the number of sink and source particles is large 
enough\footnote{The exact number required to reach peak performance depends on the used
architecture, but if the total number of gravitational interactions is $\geq 10^6$ 
it is possible to saturate the GPU} 
you can reach high performance even if the number of sinks or sources is small.

Of the two parallelisation methods the first one is most efficient when 
using a shared-time step algorithm, because fewer steps are involved in 
computing the gravity. However, the {\tt Sapporo1} method is more suitable for 
block time-stepping algorithms commonly used in high precision 
gravitational $N$-body simulations. Even though this method requires an extra step
to combine the partial results from the different subsets. The  {\tt Sapporo1} method
is also applied in this work. 
With {\tt Sapporo1} being around for 5 years we completely rewrote it 
and renamed it to {\tt Sapporo2}, which is compatible with current
hardware and is easy to tune for future generation accelerator
devices and algorithms using the supplied test scripts. 
The next set of paragraphs describe the implementation and the choices we made.

\subsection{Implementation}

\subsubsection{CUDA and OpenCL}

When NVIDIA introduced the CUDA framework in 2007 it came with 
compilers, run time libraries and examples. CUDA is an extension to the {\tt C} programming 
language and as such came with language changes. These extensions are
part of the device and, more importantly, part of the host code\footnote{The most notable 
addition is the {\tt `$<<< >>>$'} construction to start compute kernels.}. The use of 
these extensions requires that the host code is compiled using the compiler supplied 
by NVIDIA. With the introduction of the `driver API'\footnote{
	The driver API requires the use of the low-level functions formatted 
	as {\tt cuFooBar()}  while the run time API uses the higher level
functions formatted as {\tt cudaFooBar()}.}
	this was no longer required. The 
	`driver API' does not require modifications to the {\tt C} language for the host code. However, 
writing CUDA programs with the `driver API' is more involved than with the `run time API',
since actions that were previously done by the NVIDIA compiler now have to be performed by
the programmer.

When the OpenCL programming language was introduced in 2009 it came with a set of 
extensions to the {\tt C} language to be used in the device code. There are 
no changes to the language used for writing the host code, instead OpenCL comes with 
a specification of functions to interact with the device. This 
specification is very similar to the specification used in the CUDA driver API and
follows the same program flow. 

In order to support both OpenCL and CUDA in {\tt Sapporo2} we 
exploited the similarity between the CUDA driver API and the OpenCL API. We developed a 
set of {\tt C++} classes on top of these APIs which offer an unified interface for the host 
code. The classes encapsulate a subset of the OpenCL and CUDA functions for creating 
device contexts, memory buffers (including functions to copy data) and kernel operations 
(loading, compiling, launching). Then, depending on which class is included at compile time 
the code is executed using OpenCL or CUDA. The classes have no support for the more advanced
CUDA features such as OpenGL and Direct3D interoperability.

\paragraph{Kernel-code}

With the wrapper classes the host-code is language independent.
For the device code this is not the case, even though the languages are
based on similar principles, the support for advanced features like 
{\tt C++} templates, printing and debugging functionality in CUDA makes it 
much more convenient to develop in pure CUDA. Once we have a working CUDA version 
we convert this to OpenCL. The use of templates in particular
reduces the amount of code. In the CUDA version all possible 
kernel combinations are implemented using a single file with templates. 
For OpenCL a separate file has to be written for each combination of 
integrator and numerical precision. 
\\
The method used to compute the gravitational force is comparable to 
the method used in {\tt Sapporo1} with only minor changes to allow 
double precision data loads/stores and more efficient loop 
execution.

\subsubsection{Numerical Accuracy}
\label{Sapporo2:sect:numericalAccuracy}

During the development of {\tt Sapporo1} (before the GT200 chips) GPUs lacked support for IEEE-754 double precision 
computations and therefore all the compute work was done in either single or double-single precision.
The resulting force computation had similar precision as the, at that time, commonly used
GRAPE hardware~\cite{1998sssp.book.....M, Gaburov2009630}. 
This level of accuracy is sufficient for the fourth order Hermite
integration scheme~\cite{1992PASJ...44..141M,2014arXiv1402.6713P}. Currently, however
there are integrators that accurately solve the equations of motions of stars around black-holes, 
planets around stars and similar systems that encounter high mass ratios. For these kind of 
simulations one often prefers IEEE-754 double precision to solve the equations of 
motion. 
The current generation of GPUs support IEEE-754, which enables computations that
require this high level of accuracy. The data in {\tt Sapporo2}  is, therefore, always 
stored in double precision. The advantage is that we can easily add additional 
higher order integrators that require
double precision accuracy computations, without having to rewrite major parts of the 
host code. Examples of such integrators are the sixth and eighth order Hermite
integrators~\cite{2008NewA...13..498N}.
The performance impact of double precision storage on algorithms that do not require double
precision computations is limited. Before the actual computations are executed the particle 
properties are converted to either {\tt float} or {\tt double-single} and the precision therefore does 
not influence the computational performance. The penalty for loading and storing double the 
amount of data is relatively small as can be seen in the result section where {\tt Sapporo1} 
is compared to {\tt Sapporo2}.

\subsubsection{multiple GPUs}

Our new $N$-body library can distribute the computational work over 
multiple GPUs, as long as they are installed in the same system. While in {\tt Sapporo1} this
was implemented using the {\tt boost} threading library, 
this is now handled using {\tt OpenMP}. 
The multi-GPU parallelisation is achieved by parallelising over the source particles.
In {\tt Sapporo1} each GPU contained a copy of all source particles 
(as in~\cite{2007NewA...12..357H}), but
in {\tt Sapporo2} the source particles are distributed over the devices using the
round-robin method. Each GPU now only holds a subset of the source 
particles (similar to {\tt PhiGPU, HiGPU} and {\tt NBODY6}) which
reduces memory requirements, transfer time and the time to execute the 
prediction step on the source particles. However, the order of the particle distribution 
and therefore, the order in which the addition is executed is changed when comparing {\tt Sapporo1} and {\tt Sapporo2}.
This in turn can lead to differences in the least significant digit when comparing the 
computed force of {\tt Sapporo1} to {\tt Sapporo2}.

\subsubsection{Other differences}

The final difference between {\tt Sapporo1} and  {\tt Sapporo2} is the way
the partial results of the parallelisation blocks are combined. 
{\tt Sapporo1} contains two computational kernels to solve the gravitational
forces. The first computes the partial forces for the individual blocks 
of source particles, and the second sums the partial results. With the use
of atomic operators these two kernels can be combined,
which reduces the complexity of maintaining two compute kernels when 
adding new functionality, at a minimal performance impact. 
The expectation is that future devices require more active threads to saturate the GPU,
but at the same time offer improved atomic performance. The single kernel method that we 
introduced here will automatically scale to future devices and offers less overhead than
launching a separate reduction kernel. This reduced overhead results in slightly 
better performance (few \%) on current architectures compared to the original 
two kernel method. 
In total we now require three GPU kernels to compute gravity, 
one copy kernel to move particles from CPU buffers to GPU buffers,
one kernel to predict 
the particles to the new time-step and finally, the gravity kernel to compute 
the forces.

\section{Results}
\label{Sapporo2:Sect:Results}

In astrophysics the current most commonly used integration method is the 
fourth order Hermite~\cite{1992PASJ...44..141M} integrator. This integrator
requires the velocity, the acceleration and the first time derivative of the acceleration (jerk) 
to be computed. The integrator furthermore requires information of the nearest 
neighbouring particle, this to determine collisional events or binary formation. 
Finally, the more advanced integrators such as {\tt NBODY4}~\cite{1999PASP..111.1333A}
and Kira~\cite{2001MNRAS.321..199P} require a list of particles within a given radius from
each particle to determine the perturber list. All this is what 
{\tt Sapporo1} computes and how the GRAPE hardware operates~\cite{1998sssp.book.....M}. 
The used numerical precision in this method is the double-single variant. 
In order to compare the new implementation with the results of {\tt Sapporo1}, all
results in this section, unless indicated otherwise, refer to the double-single fourth 
order Hermite integrator. Furthermore, we have enabled the computation of 
the nearest neighbour and the list of nearby particles, as has {\tt Sapporo1}.
However if the user does not require this information 
it can be disabled by changing a template parameter in the code. 

For the performance tests we used different machines, depending on which GPU was used.
All the machines with NVIDIA GPUs have {\tt CUDA 5.5} toolkit and drivers installed.
For the machine with the AMD card we used version 2.8.1.0 of the APP-SDK toolkit and driver version 13.4.

The full list of used GPUs can be found in Tab.~\ref{Sapporo2:Tab:GPUs}, the table shows  
properties such as clock speed and number of cores. In order to compare the various GPUs
we also show the theoretical performance, relative with respect to the {\tt GTX480}. Since,
theoretical performance is not always reachable we also show the relative practical
performance as computed with a simple single precision $N$-body kernel that is designed for 
shared-time steps, similar to the $N$-body example in the CUDA SDK~\cite{Nyland_nbody}. 

\subsection{Thread-block configuration}
\label{Sapporo2:sect:tbc}

{\tt Sapporo2} is designed around the concept of processing a fixed number of 
sink particles for a block time-step algorithm (see Section~\ref{Sapporo2:Sect:Method:Par}). 
Therefore, the first thing to determine is the smallest number of sink particles that gives full GPU performance.
To achieve full performance the computation units on the GPUs have to be saturated with work.
The GPU consists of a number of multiprocessors and the computation units are spread over these multiprocessors.
When the host code sends work to the GPU this is done in sets of thread-blocks. Each thread-block is executed 
on a multiprocessor. The blocks contain a (configurable) number of threads that can work together, while the blocks
themselves are treated as independent units of work.
In this section we determine the optimal number of blocks and the number of threads per block to saturate the GPU
when performing the gravity computations. 
We test a range of configurations where we vary the number of blocks per multi-processor
and the number of threads per block. The results for four different GPU architectures
are presented in Fig.~\ref{Sapporo2:fig:BlockAndThreadConfigs}. In this figure each line 
represents a certain number of blocks per multi-processor, $N_{blocks}$. The x-axis 
indicates the number of threads in a thread-block, $N_\text{threads}$. The range of this
axis depends on the hardware. For the {\tt HD7970} architecture we cannot 
launch more than $N_\text{threads}=256$, and for the {\tt GTX480} the limit is 
$N_\text{threads}=576$. For the two {\tt Kepler} devices {\tt 680GTX} and {\tt K20m} we can launch
up to $N_\text{threads}=1024$ giving these last two devices the largest set of configuration options.
The y-axis shows the required wall-clock time to compute the forces using the 
indicated configuration, the bottom line indicates the most optimal configuration.  

For the {\tt 680GTX} and the {\tt K20m} the $N_\text{blocks}$ configurations 
reach similar performance when  $N_\text{threads} > 512$.  This indicates
that at that point there are so many active threads per multi-processor, that there 
are not enough resources (registers and/or shared-memory) to accommodate multiple thread-blocks 
per multi-processor at the same time. To make the code suitable for 
block time-steps the configuration with the least number of threads, that gives 
the highest performance would be the most ideal. For the {\tt HD7970} this 
is $N_\text{threads}=256$ while for the {\tt Kepler} architectures $N_\text{threads}=512$ 
gives a slightly lower execution time than $N_\text{threads}=256$ and $N_\text{threads}=1024$.
However, we chose to use $N_\text{threads}=256$ for all configurations and
use 2D thread-blocks on the {\tt Kepler} devices to launch 512 or 1024 threads. 
When we talk about 2D thread-blocks it means that we launch multiple threads
per $i$-particle whereby each thread computes a part of the $j$-particles. 
This way we increase the number of total threads which the hardware can 
schedule in order to hide the memory latencies. Especially when the number 
of active $i$ particles is $\leq 128$ this helps to improve 
the performance and is discussed in more detail in the next section.
For each architecture the default configuration is indicated with the 
circles in Fig.~\ref{Sapporo2:fig:BlockAndThreadConfigs}.

\subsection{Block-size / active-particles}

Now we inspect the performance of {\tt Sapporo2} in combination with a block time-step algorithm.
We measured the time to compute the gravitational forces using either the 
NVIDIA GPU Profiler or the built-in event timings of OpenCL. 
The number of active sink particles, $N_\text{active}$, is varied between 1 and 
the optimal $N_\text{threads}$ as specified in the previous paragraph. 
The results are averaged over 100 runs and presented in Fig.~\ref{Sapporo2:fig:threadPerformance}.
We used 131072 source particles which is enough to saturate the GPU
and is currently the average number of particles used in direct $N$-body simulations
that employ a block time-step integration method.

The straight striped lines in Fig.~\ref{Sapporo2:fig:threadPerformance} indicate the theoretical linear scaling from 
$(0,0)$ to $(256, X)$ where $X$ is the execution time of the indicated GPU
when $N_\text{active}=256$. Visible in the figure are the jumps in the 
execution time that coincide with the warp (wavefront) size of 32 (64). 
For NVIDIA devices we can start 2D thread-blocks for all values of $N_\text{active}$, 
since the maximum number of threads that can be active on the device is $\geq 512$. 
The effect of this is visible in the more responsive execution times
of the NVIDIA devices when decreasing $N_\text{active}$ compared to the AMD device. Each time 
$N_\text{active}$ drops below a multiple of the maximum number of active threads,
the execution time will also decrease. When $N_\text{active}$ decreases from 
 $N_\text{active} \aplt 64$  the execution
time goes down linearly, because of the multiple blocks that can be 
started for any value of $N_\text{active}$. The lines indicated with `1D' in the 
legend show the execution time, if we would not subdivide the work further using 
2D thread-blocks. This will under-utilize the GPU and results in increased execution
times for $N_\text{active} < 128$.

The performance difference between {\tt CUDA} and  {\tt OpenCL} is minimal, 
which indicates that the compute part of both implementations inhabits similar 
behavior. 
For most values of $N_\text{active}$ the timings of {\tt Sapporo1} and {\tt Sapporo2} 
are comparable. Only for $N_\text{active} < 64$ we see a slight advantage for {\tt Sapporo1}
where the larger data loads of {\tt Sapporo2} result in a slightly longer execution time.
However, the improvements made in {\tt Sapporo2} result in higher performance and 
a more responsive execution time compared to {\tt Sapporo1} when $128 \geq  N_\text{active} < 160$.
For the {\tt HD7970}, there is barely any improvement when $N_\text{active}$ decreases from 256 to 128.
There is a slight drop in the execution time at $N_\text{active}=192$, which
coincides with one less active wavefront compared to $N_\text{active}=256$.
When $N_\text{active} \leq 128$ we can launch 2D blocks and the performance improves again
and approaches that of the NVIDIA hardware, but the larger wavefront size compared to 
the warp size causes the the execution times to be less responsive to changes of $N_\text{active}$.

\subsection{Range of N}
\label{Sapporo2:Sect:RangeOfN}

Now that we selected the thread-block configuration we continue with 
testing the performance when computing the gravitational forces 
using $N_\text{sink}$  and $N_\text{source}$ particles, resulting in 
$N_\text{sink} \times N_\text{source}$
force computations (we set $N_\text{sink} =  N_\text{source}$).
The results are presented in the left panel of
Fig.~\ref{Sapporo2:fig:NxNPerformance}. This figure shows the results for the 
five GPUs using {\tt CUDA, OpenCL, Sapporo1} and {\tt Sapporo2}.
The execution time includes the time required to send the input 
data and retrieve the results from the device.

The difference between {\tt Sapporo1} and {\tt Sapporo2} (both the {\tt CUDA} 
and {\tt OpenCL} versions) on the K20m GPU are negligible. {\tt Sapporo1} is
slightly faster for $N < 10^4$, because of the increased data-transfer 
sizes in {\tt Sapporo2}, which influence the performance more when the number of
computations is relatively small. 
{\tt Sapporo2} is slightly faster than  {\tt Sapporo1} when $N \geq 10^4$,
because of the various optimisations added to the new version. 
The difference between the 
{\tt GTX680}, {\tt K20m} and the {\tt HD7970} configurations is relatively small. 
While the {\tt GTX Titan} is almost $1.5\times$ faster and the {\tt GTX480} 
almost $2\times$ slower than these three cards. These numbers are not unexpected
when inspecting their theoretical performance (see Tab.~\ref{Sapporo2:Tab:GPUs}). 
For $N < 10^5$ we further see that the 
performance of the {\tt HD7970} is lower than for the NVIDIA cards. This difference 
is caused by slower data transfer rates between the host and device for the {\tt HD7970}. 
Something similar can be seen when we compare the {\tt OpenCL} version of the {\tt K20m} with the 
{\tt CUDA} version. 
Close inspection of the timings indicate that this difference is caused by
longer CPU-GPU transfer times in the {\tt OpenCL} version when transferring small 
amounts of data ($< 100$KB) which, for small $N$, forms a larger part of the total 
execution time.

\subsection{Double precision vs Double-single precision}

As mentioned in Section~\ref{Sapporo2:sect:numericalAccuracy} the higher order integrators
require the use of double precision computations. Therefore, 
we test the performance impact when using full native 
double precision instead of double-single precision. 
For this test we use the {\tt GTX680}, {\tt K20m} and the {\tt HD7970}. 
The theoretical peak performance when using double precision computations is
lower than the peak performance when using single precision computations. 
The double precision performance of the {\tt K20m} is one third that 
of the single precision performance. For the  {\tt GTX680} this is $1\over24$th
and for the {\tt HD7970} this is one fourth. 
As in the previous section we use the wall-clock time required to 
perform $N^2$ force computations (including the data send and receive time)
to compare the devices. The results are presented in the 
right panel of Fig.~\ref{Sapporo2:fig:NxNPerformance}, here the double precision 
timings are indicated with the open symbols and the double-single timings 
with the filled symbols. 

As in the previous paragraph, when using double-single precision 
the performance is comparable for all three devices.
However, when using double-precision the differences  
become more clear. As expected, based on the theoretical numbers, the 
{\tt GTX680} is slower than the other two devices. 
The performance of the {\tt K20m} and the {\tt HD7970} are comparable for $N > 10^4$.
For smaller $N$ the performance is more influenced by the transfer rates between 
the host and the device than by its actual compute speed.

Taking a closer look at the differences we see that the performance of the 
{\tt GTX680} in full double precision is about $\sim10\times$ lower than when 
using double-single precision. For the other two cards the double precision 
performance is roughly $\sim2.8\times$ lower.
For all the devices this is roughly a factor of 2 difference from what can be
expected based on the specifications. This difference can be explained by the 
knowledge that the number of operations is not exactly the same for the two 
versions\footnote{Double-single requires more computations than
single precision on which the theoretical numbers are based} and even in the double 
single method we use the special operation units to compute the {\tt rsqrt}\footnote{An 
optimized function that computes the reciprocal-square-root ($1 / \sqrt{x}$).}. 
Another reason for the discrepancy between the practical and theoretical numbers is 
that we keep track of the nearest neighbours which requires the same operations for the
double-single and the double precision implementation. 
Combining this with the knowledge that we 
already execute a number of double precision operations to perform atomic 
additions and data reads, results in the observed difference between the 
theoretical and empirically found performance numbers.

\subsection{Sixth order performance}

The reason to use sixth order integrators compared to lower order integrators is that, 
on average, they are able to take larger time-steps. They are also better in handling systems 
that contain large mass ratios (for example when the system contains a supermassive black-hole).
The larger time-step results in more active particles per block-step which improves the GPU efficiency. 
However, it also requires  more operations than a fourth order integrator, something 
which is discussed in detail in~\cite{2008NewA...13..498N}. 
Previous work~\cite{keigo_thesis, 2013JCoPh.236..580C, 2013CoPhC.184.2528C}
indicates that double-single accuracy is sufficient for a sixth order integrator. However,
to give the user the choice we implemented both a double-single and a double precision
version of this method.
The performance results of these versions are presented in Fig.~\ref{Sapporo2:fig:4thvs6h}. 
As in the previous figures we present the time to 
compute $N^2$ forces. 
Presented are the performance of the sixth and fourth order kernels using double precision 
and using double-single precision. As expected, the sixth order requires more time  
than the fourth order as it executes the most operations.
The difference between the fourth order in double-single precision and the sixth order in 
double-single precision is about a factor 2. When we use double precision instead of 
double-single precision for the sixth order method then the execution time goes up 
by another factor of 2. The difference between the double precision fourth order and the 
double precision sixth order is about a factor of 1.4. 
The factor 2 difference in performance is relatively small and expected from the operation 
count. Therefore, if the sixth order allows you to take time-step that are two or more times larger
than when using a fourth order your total execution time will go down when using a sixth 
order integrator. This combined with the benefits of the sixth order integrator such as 
being able to integrate high mass ratios, where high accuracy is required to trace tight orbits,
makes the sixth order method a viable solution for $N$-body methods.

\subsection{Multi-GPU}

As described in Section~\ref{Sapporo2:Sect:Method}, {\tt Sapporo2} supports multiple
GPUs in parallel. The parallelised parts are the force computation, 
data transfer and prediction of the source particles. The transfer of 
particle properties to the device and the transfer of the  force computation 
results from the device are serial operations. 
These operations have a small but constant overhead, independent of the 
number of GPUs.
For the measurements in this section we use the total wall-clock
time required to compute the forces on $N$ particles (as in Section~\ref{Sapporo2:Sect:RangeOfN}).
The speed-up compared to 1 GPU is presented in Fig.~\ref{Sapporo2:fig:multiGPUPerformance}.
The timings are from the {\tt K20m} GPUs which have enough memory to store
up to $8\times10^6$ particles. We use shared-time steps for these timings. 
For $N > 10^4$ it is efficient to use all available GPUs in the system
and for $N \leq 10^4$ all multi-GPU configurations show similar performance. 
The only exception here is when $N = 10^3$ at which point the overhead of 
using 4 GPUs is larger than the gain in compute power. 
For large enough $N$ the 
scaling is near perfect ($T_\text{single-GPU}/T_\text{multi-GPU}$),
since the execution time is dominated by the computation
of the gravitational interactions. Note that for these experiments we have to transfer
the full data-sets to the GPU, this is why the scaling for small $N$ is less than perfect
as it takes time to transfer the data over a PCI-Express bus. 
For block time-step simulations the number of particles being transferred, per time-step, will be 
smaller. However, the compute time is also smaller as less particles will have to integrated.
Therefore, the scaling for small $N$ will stay less than perfect in all situations.

\subsection{Block time-step simulations}

To test the performance of the multi-GPU implementation for
block time-step simulations with {\tt Sapporo2} we use a sixth order Hermite integrator 
with block time-steps~\cite{2012ApJ...753...85F, 2008NewA...13..498N}.
We perform simulations of Plummer~\cite{1915MNRAS..76..107P} spheres using 1 
and 4 GPUs with double-single (DS) and full double precision (DP) accuracy. The number of 
particles used ranges from 16k up to 512k particles. For each simulation 
we record the execution time, the energy error, the average number of active 
particles per block-step and the speed-up of using 4 GPUs over 1 GPU.

The chosen time-step criteria is critical when performing block time-step simulations.
For fourth order Hermite the method most commonly used is the Aarseth 
method~\cite{2003gnbs.book.....A}. 
For the sixth order a generalized version of the Aarseth criterion can 
be used as, described in~\cite{2008NewA...13..498N}.
However, this generalized version is unstable when the force computation is not accurate
enough\footnote{Keigo Nitadori \& Michiko Fujii, private communication.}.
Specifically, rounding errors in the jerk and snap computation can cause the 
time-step to go to zero. Before running production simulations one should 
carefully consider which accuracy and time-step method to use, however a full analysis of 
the best time-step method for these situations is beyond the scope of this work.  
In \cite{2015arXiv150101040S} the authors work
around this time-step problem by taking the average of the Aarseth fourth order method and the sixth order  
extension to compute the time-step (their Eq.~8). In order to compare
the timing and accuracy of our simulations 
we use this average method for both our 
DS and DP simulations. Note that using the sixth order 
time-step computation together with DS force computation may result
in a time-step that approaches zero. While the sixth order time-step combined with 
full DP force computation will work without problems.

For these simulations we set $\eta_4=0.01$ and $\eta_6=0.1$ and simulate 
the model for one $N$-body time-unit. The presented execution times cover the full execution
from the start to the end of a simulation. The time therefore, includes all required
operations on the GPU side (predict, gravity, particle copy) as well as on the host side
(corrections, time-step computation, particle copies). 
During the simulation the size of $N_\text{active}$ varies between 1 and $N$.

The resulting data for the simulations are presented in Fig.~\ref{Sapporo2:fig:hermite6}. 
The figure contains four panels, the top left panel presents the absolute
execution time. The top right panel the speed-up when scaling from 1 to 4 GPUs.
The bottom left panel the average number of particles that is being
integrated, $N_\text{active}$. Finally, the bottom right panel presents
the energy error at the end of the simulation.
For all panels the solid lines indicate the simulations that use a single GPU
and the dashed lines indicate the simulations with four GPUs. The square symbols
indicate the simulations that use DS accuracy and the DP runs are 
indicated by the round symbols.

The execution time scales, as expected, as $O(N^2)$ and as we can see 
in the bottom left panel that the average number of active particles 
increases with the total number of particles.

There are a number of other things we can see in the figures. First of all
we can see that the full double precision simulations run faster than
the double-single simulations. Eventhough the compute work is faster for the 
double-single version (as we saw in Fig.~\ref{Sapporo2:fig:multiGPUPerformance}),
the reduced accuracy forces 
the integrator to take more smaller time-steps. This can be seen by the 
average number of particles per block
which is smaller for the DS simulations than for the DP simulations.
Another thing to note is that the results of the single GPU DS simulations are 
slightly different than the four GPU DS simulations. This is another consequence 
of the reduced accuracy, the changed addition order when running on more than
a single GPU results in rounding differences. For DP the results for single 
and multi GPU simulations are so similar that the differences are not visible in the figures.
The DP simulations are not only faster, they also produce an enery error
that is almost two orders of magnitude smaller than that of the DS 
simulations. The energy error for the DP simulations is around 
$10^{-12}$ and that of the DS simulations around $10^{-10}$.

In Fig.~\ref{Sapporo2:fig:multiGPUPerformance} we saw that the speed-up when going from 1 to 4 GPUs 
scales from a factor 1 to 4x when the number of particles increases. 
A similar effect we see occuring in the bottom right panel; when the number of active 
particles increases the speed-up also increases.
The jump in speed-up for the DS when going from
256k particles to 512k particles is caused by the increase of 
$N_\text{active}$ between 256k and 512k.

These simulations show that the benefit of using more than a single GPU depends on the dataset 
size, the used accuracy as well as on the average size of $N_\text{active}$. It is therefore
important that one knows these numbers when performing many simulations.
Especially, when using a sixth order integrator, as we did here, it is critical that one
chooses a time-step method that is suitable for the used accuracy.

\section{Discussion and CPU support}

\subsection{CPU}

With the availability of CPUs with 8 or more cores that support advanced vector instructions
there is the recurring question if it is not faster to compute the gravity on the CPU 
than on the GPU. Especially since there is no need to transfer data between the host 
and the device, an operation which 
can be relatively costly when the number of particles is $\leq1024$. To test exactly for 
which number of particles the CPU is faster than the GPU we added a CPU implementation to
{\tt Sapporo2}. This CPU version uses SSE2 vector instructions and OpenMP 
parallelisation and can be run in single or in double precision. 
The only kernel implemented is the fourth order integrator, including 
support for neighbour lists and nearest neighbours (particle-ID and distance). 
Because the performance of the GPU depends on the combination of sink and source
particles we test a grid of combinations for the number of sink and source particles
when measuring the time to compute the gravitational forces.
The results for the CPU (a Xeon E5620 @ 2.4Ghz), using a single core, are presented in Fig.~\ref{Sapporo2:fig:CPUvsGPU}a. 
In this figure (and all the following figures) the x-axis indicates the number of sinks 
and the y-axis the number of sources. 
The execution time is indicated by the colour from blue (fastest) to red (slowest). 
The smooth transition from blue to red from the bottom left corner
to the top right indicates that the performance does not preferentially depend on either 
the source or sink particles, but rather on the combined number of interactions.
This matches our expectations, because the parallelisation granularity on the CPU is as 
small as the vector width, which is 4. 
On the GPU this granularity is much higher, as presented in Fig.~\ref{Sapporo2:fig:CPUvsGPU}b,
here we see bands of different colour every 256 particles. Which corresponds to
the number of threads used in a thread-block ($N_\text{threads}$).
With 256 sink particles we have the most optimal
performance of a block, however, if we would have 257 sink particles we process the first 256 
sinks using optimal settings while the 257th sink particle is processed relatively inefficiently. 
This granularity becomes less obvious when we increase the number of interactions 
as presented in Fig.~\ref{Sapporo2:fig:CPUvsGPU}c. Here we see the same effect appearing as with
the CPU (Fig.~\ref{Sapporo2:fig:CPUvsGPU}a), where the granularity becomes less visible once we 
saturate the device and use completely filled thread-blocks for most of the particles.
The final panel, Fig.~\ref{Sapporo2:fig:CPUvsGPU}d, indicates per combination of source 
and sink particles which CPU or GPU configuration is the fastest.
For the CPU we measured the execution time when using 1,2,4 or 8 cores. In this panel
the colours indicate the method which gives the shortest execution times. Furthermore
does it indicate if and by how much the GPU is faster than the 8 cores of the CPU.

When either the number of sinks or the number of sources is relative small 
($\leq 100$) the CPU implementation performs best. 
However, when the number of sinks or sources is $>100$ the 
GPU outperforms the CPU. When using a CPU implementation that uses the AVX or AVX2 
instruction sets the borders of these regions would shift slightly upwards. The 
CPU would then be faster for a larger number of source/sink particles, but that would 
only be at most for a factor of 2 to 4 more particles. The data of Fig.~\ref{Sapporo2:fig:CPUvsGPU}
confirms that our choice to implement the {\tt Sapporo2} library for the GPU is an 
efficient method for realistic data-set sizes. 
Although our implementation uses SSE2 instructions it is not as advanced as the implementation
of~\cite{2012NewA...17...82T}. For example, we use intrinsic functions while they use the assembly operations directly.
This is also visible when we compare their performance with our implementation. The implementation we tested here reaches
about 60\% of their performance, however they do not compute the nearest neighbour particle 
and do not keep track of the neighbourlist, both of which have a significant impact on the performance 
as they cause divergence in the execution stream.

\subsection{XeonPhi}

Because the {\tt Sapporo2} library can be built with {\tt OpenCL} it should, theoretically, be 
possible to run on any device that supports {\tt OpenCL}. To put this to the test, we compiled
the library with the Intel {\tt OpenCL} implementation.
However, although the code compiled without problems it did not produce correct results.
We tested the library both on an Intel CPU and the Intel {\tt XeonPhi} accelerator. Neither the CPU,
nor the {\tt XeonPhi} produced correct results. Furthermore, the performance 
of the {\tt XeonPhi} was about 100$\times$ smaller than what can be expected from its theoretical
peak performance. We made some changes to the configuration parameters such as
$N_\text{threads}$ and $N_\text{blocks}$, however this did not result in any presentable performance.
We suspect that the Intel {\tt OpenCL} implementation, especially for XeonPhi, contains a number of limitations that
causes it to generate bad performing and/or incorrect code. Therefore, the {\tt Sapporo2} library
is not portable to Intel architectures with their current {\tt OpenCL} implementation\footnote{
	A short test on an AMD CPU gave correct results therefore we suspect it is something intrinsic to
the Intel {\tt OpenCL} environment}.
This does not imply that the {\tt XeonPhi} has bad performance in general, since it is possible to
achieve good performance on $N$-body codes that is comparable to GPUs. However, this requires code
that is specifically tuned to the {\tt XeonPhi} architecture (K. Nitadori, private 
communication~\footnote{Also see {\tt https://github.com/nitadori/Hermite} and \\
{\tt http://research.colfaxinternational.com/post/2013/01/07/Nbody-Xeon-Phi.aspx}.}).

\section{Conclusion}

The here presented {\tt Sapporo2} library makes it easy to enable GPU acceleration 
for direct $N$-body codes. We have seen that the difference between the 
{\tt CUDA} and {\tt OpenCL} implementation is minimal, when there are enough 
particles to make the simulation compute limited. However, if many small data 
transfers are required, for example when the integrator takes very small time-steps
with few active particles, the {\tt CUDA} implementation will be faster.
Apart from the here presented fourth and sixth order integrators the library 
also contains a second order implementation. And because of the storage of data 
in double precision it can be trivially expanded with an eighth order integrator. 
The performance gain when using multiple GPUs implies that it is efficient to 
configure GPU machines that contain more than 1 GPU. This will improve the time 
to solution for simulations with more than $10^4$ particles. 

The {\tt OpenCL} support and built-in tuning methods allow easy extension to 
other {\tt OpenCL} supported devices. However, this would require a mature {\tt OpenCL} 
library and matching hardware that supports atomic operations and double precision data types. For the {\tt CUDA}
devices this is not a problem since the current {\tt CUDA} libraries already have 
mature support for the used operations and we expect that the library automatically scales to future architectures. 
The only property that has to be set is the number of thread-blocks per multiprocessor 
and this can be easily identified using the figures as presented in Section~\ref{Sapporo2:sect:tbc}.

The library is freely available either as part of the AMUSE software 
package~\cite{2013CoPhC.184..456P},  which can be downloaded
from: http://wwww.amusecode.org. or as standalone library from: https://github.com/treecode/sapporo2/.




\begin{backmatter}

\theendnotes

\section*{Competing interests}
  The authors declare that they have no competing interests.


\section*{Acknowledgements}

We thank the anonymous reviewers for their extensive and helpful comments.
This work was supported
by the Netherlands Research Council NWO (grants \#643.200.503, 
\# 639.073.803, \#614.061.608, \# 612.071.503, \#643.000.802).


\bibliographystyle{bmc-mathphys} 
\bibliography{sapporo2}      




\section*{Figures}

\begin{figure}[h!]
  \centering
  \includegraphics[width=0.95\columnwidth,natwidth=360,natheight=360]{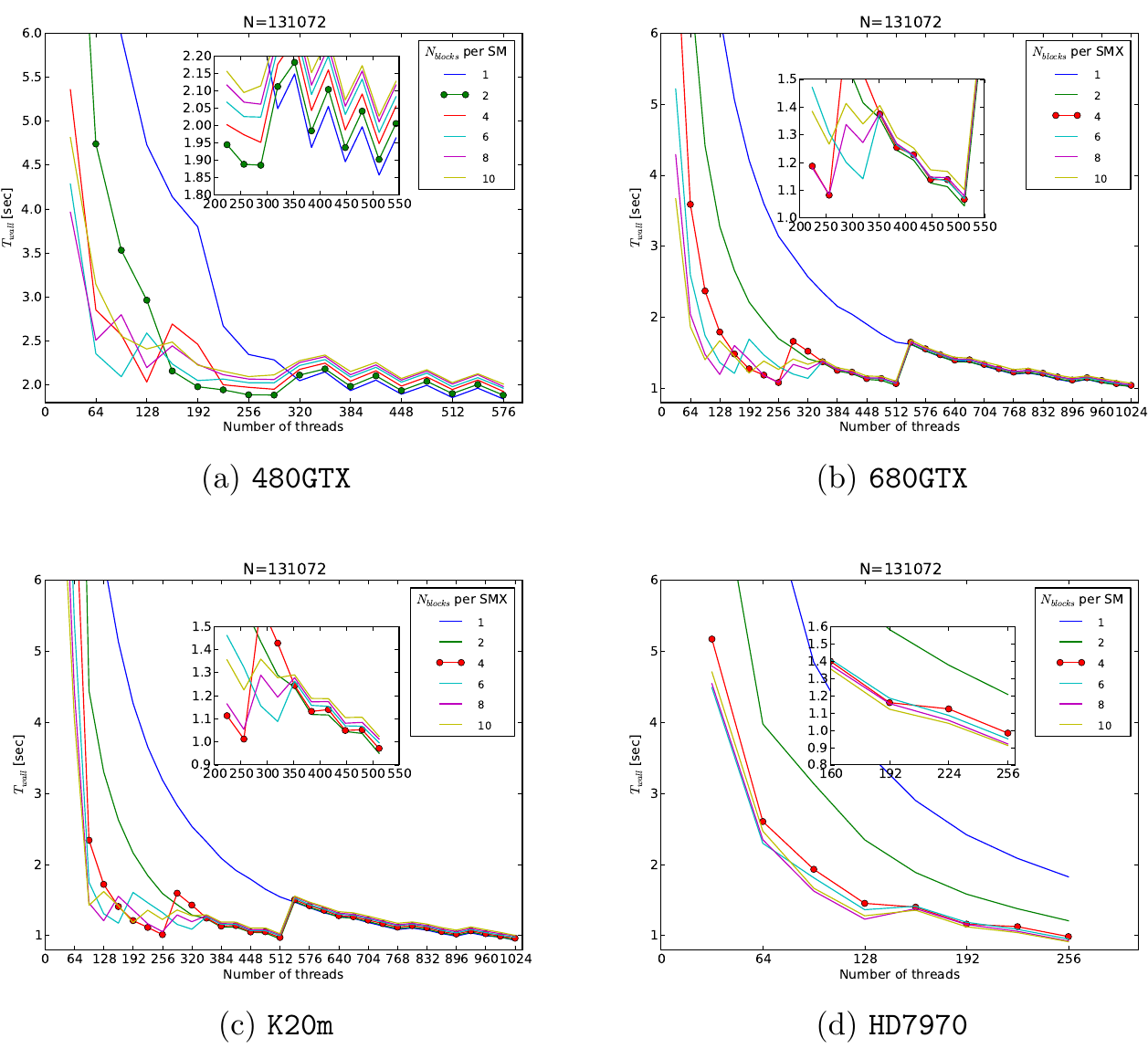}
  \caption{Performance for different thread-block configurations. The figure shows the 
required integration time (y-axis)  for $N=131072$ source particles using different
number of sink particles (number of threads, x-axis). Each line indicates a different
configuration. In each configuration we changed the number of blocks launched
per GPU multi-processor for different GPU architectures. Shown in panel (a) 
NVIDIA's  {\tt Fermi} architecture, in panel (b) the NVIDIA {\tt Kepler, GK104} architecture
in panel (c) the NVIDIA {\tt Kepler, GK110} and the AMD {\tt Tahiti} architecture
in panel (d). The AMD architectures are limited to 256 threads.
The configurations that we have chosen as our default settings for the number of 
blocks are the lines with the filled circle markers.}
\label{Sapporo2:fig:BlockAndThreadConfigs}
\end{figure}

\begin{figure}[h!]
\includegraphics[width=\columnwidth]{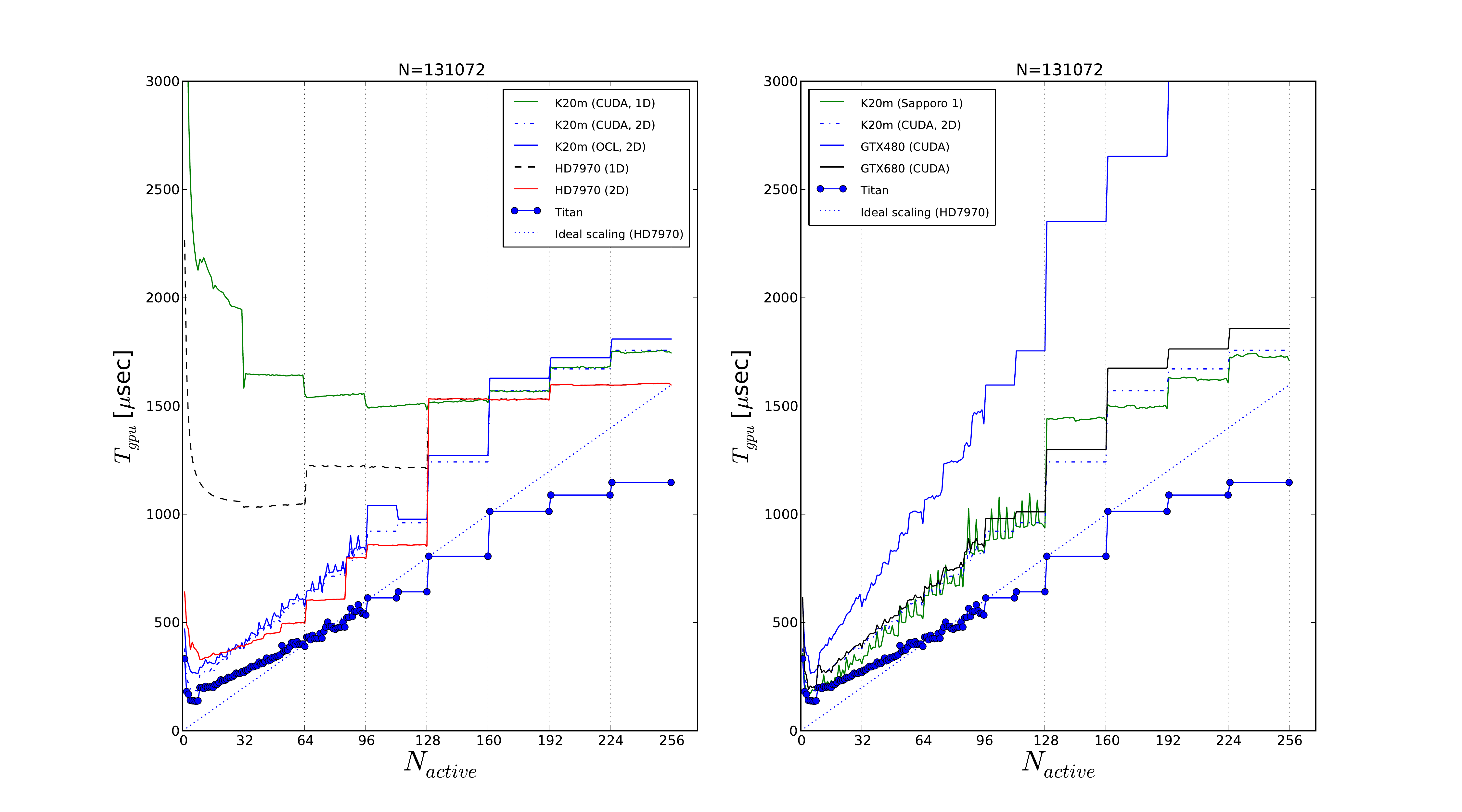}
\caption{Performance for different numbers of active sink particles. The x-axis  
indicates the number of active particles and the y-axis the required time
to compute the gravitational force using 131072 source particles 
($N_{active} \times N$ gravity computations). The presented time only includes the time
required to compute the gravity, the data transfer times are not included.
In both panels the linear striped line shows the ideal scaling from the most 
optimal configuration with 256 active particles to the worst case situation
with 1 active particle for one of the shown devices.
The left panel shows the effect on the performance when using 1D thread-blocks
instead of 2D on AMD and NVIDIA hardware. It also we shows the effect
of using {\tt OpenCL} instead of {\tt CUDA} on NVIDIA hardware. When using 1D
thread-blocks the GPU becomes underutilized when $N_{active}$ becomes 
smaller than $\sim128$. This is visible as the execution time increases
while $N_{active}$ becomes smaller. 
The right panel compares the performance of the five different GPUs as indicated. 
Furthermore, it shows that the performance of {\tt Sapporo2} is 
comparable to that of {\tt Sapporo1}. }
\label{Sapporo2:fig:threadPerformance}
\end{figure}

\begin{figure}[h!]
\includegraphics[width=\columnwidth]{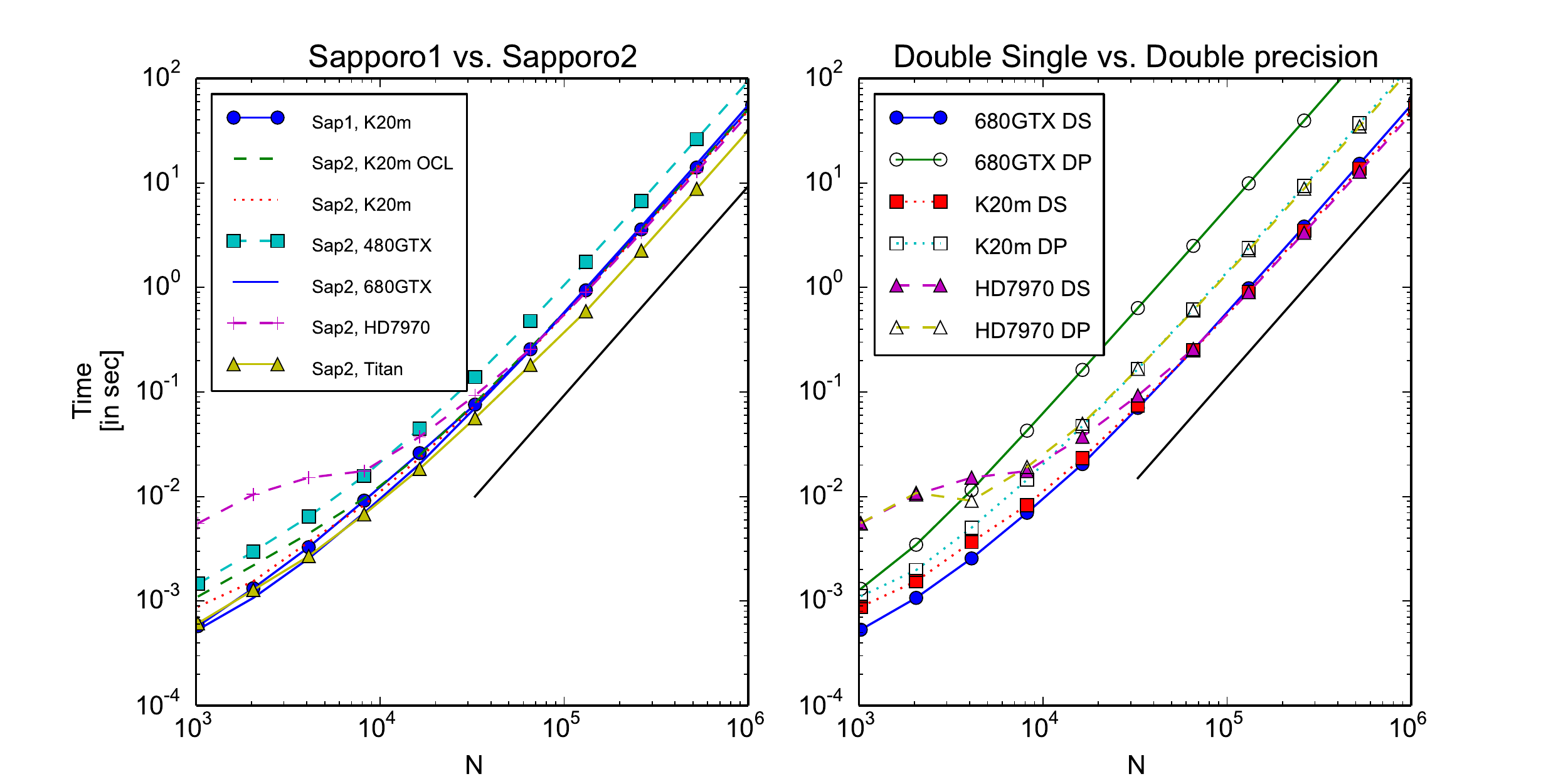}
\caption{Time required to solve $N^2$ force computations using different
configurations. In both panels the number of source particles is equal to 
the number of sink particles, which is indicated on the x-axis. The y-axis indicates
the required wall-clock time to execute the gravity computation and 
to perform the data transfers. Unless 
otherwise indicated we use CUDA for the NVIDIA devices.
The left panel shows the performance of {\tt Sapporo1} on a {\tt K20m} GPU
and {\tt Sapporo2} on 5 different GPUs using a mixture of  {\tt CUDA} and
{\tt OpenCL}. The straight solid line indicates $N^2$ scaling. 
The right panel shows the difference in performance between double-single
and double precision. We show the performance for three different devices. The
double-single timings are indicated by the filled symbols. The double-precision
performance numbers are indicated by the lines with the open symbols. 
The straight solid line indicates $N^2$ scaling. }
\label{Sapporo2:fig:NxNPerformance}
\end{figure}

\begin{figure}
\includegraphics[width=\columnwidth]{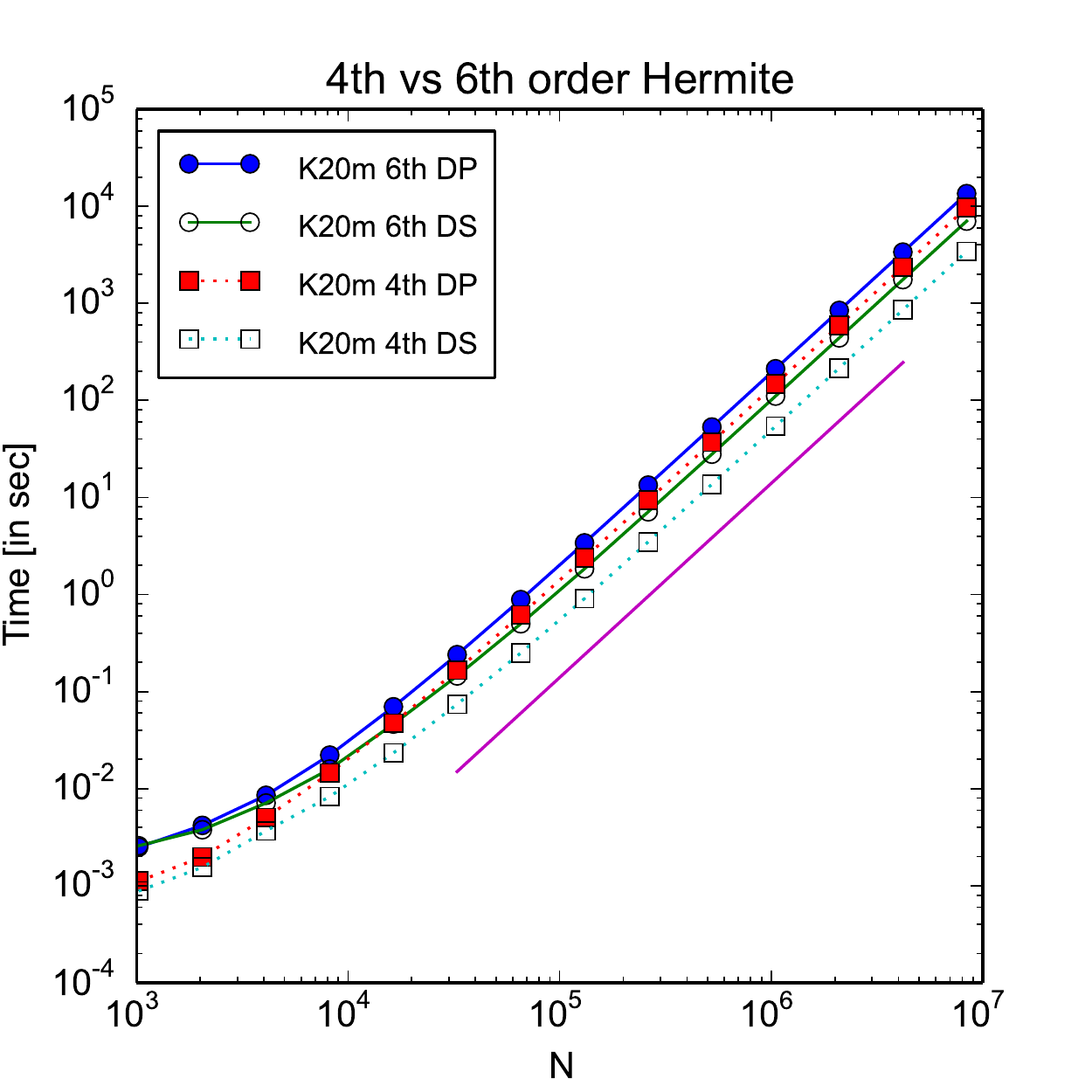}
\caption{Performance difference between fourth and sixth order kernels.
Shown is the time required to solve $N^2$ force computations using different
configurations. The number of source particles is equal to 
the number of sink particles indicated on the x-axis. The y-axis indicates
the required wall-clock time to execute the gravity computation 
and to perform the data transfers. The fourth-order configuration using 
double-single precision is indicated by the dotted line with open square symbols.
The fourth order configuration using double precision is indicated by the 
dotted line with filled square symbols. The sixth order configuration 
using double-single precision is indicated by the solid line with open circles 
and the sixth order with double precision is indicated by the
solid line with filled circles. 
The straight solid line without symbols indicates the $N^2$ scaling.
Timings performed on a {\tt K20m} GPU using {\tt CUDA 5.5}.}
\label{Sapporo2:fig:4thvs6h}
\end{figure}

\begin{figure}[h!]
\includegraphics[width=\columnwidth]{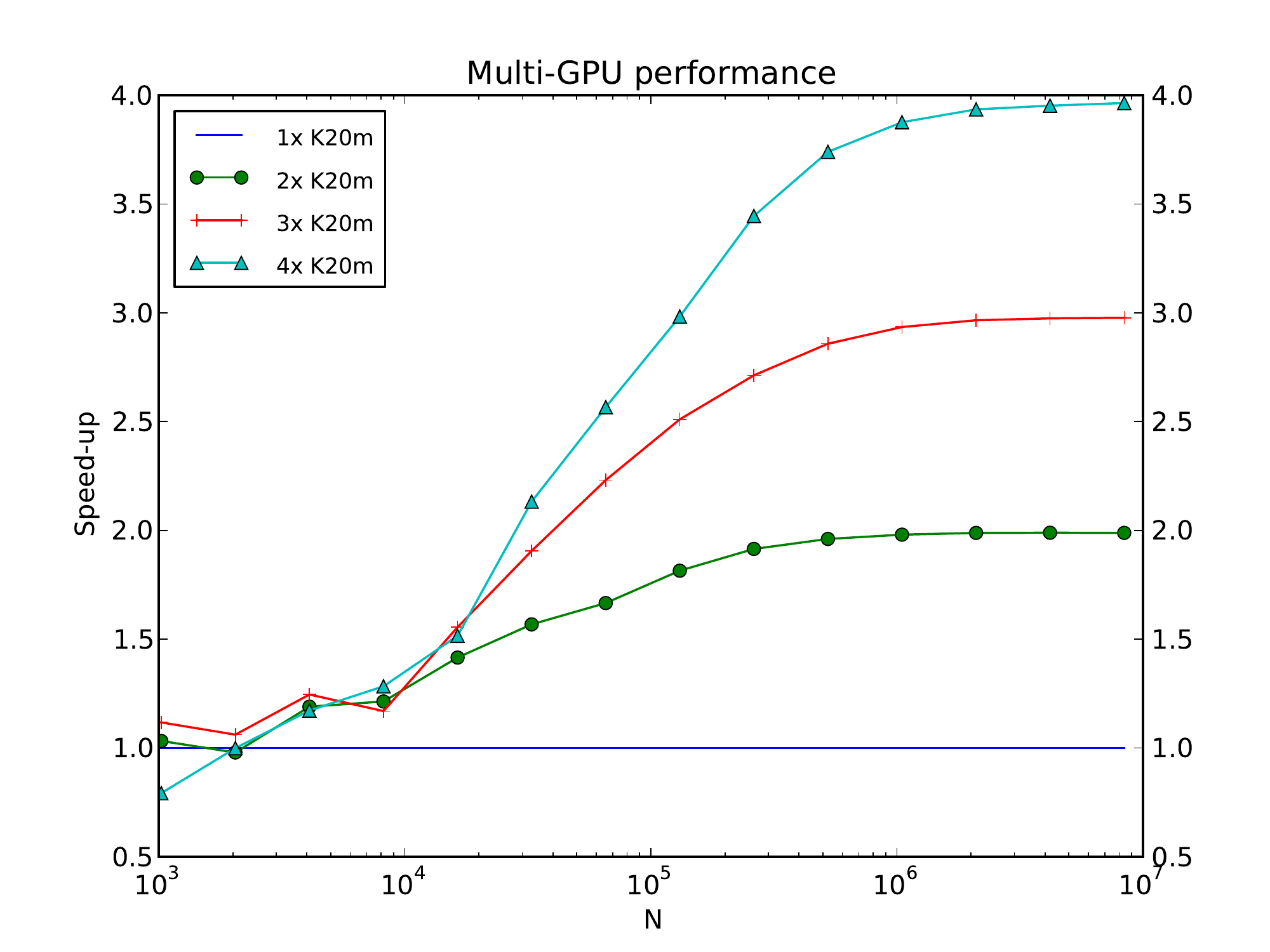}
\caption{Multi-GPU speed-up over using one GPU. For each configuration 
the total wall-clock time is used to compute the speed-up (y-axis) 
for a given $N$ (x-axis). The wall-clock time includes the time 
required for the reduction steps and data transfers. 
Timings performed on {\tt K20m} GPUs using {\tt Sapporo2} and  
{\tt CUDA 5.5}.}
\label{Sapporo2:fig:multiGPUPerformance}
\end{figure}

\begin{figure}
\includegraphics[width=0.95\columnwidth]{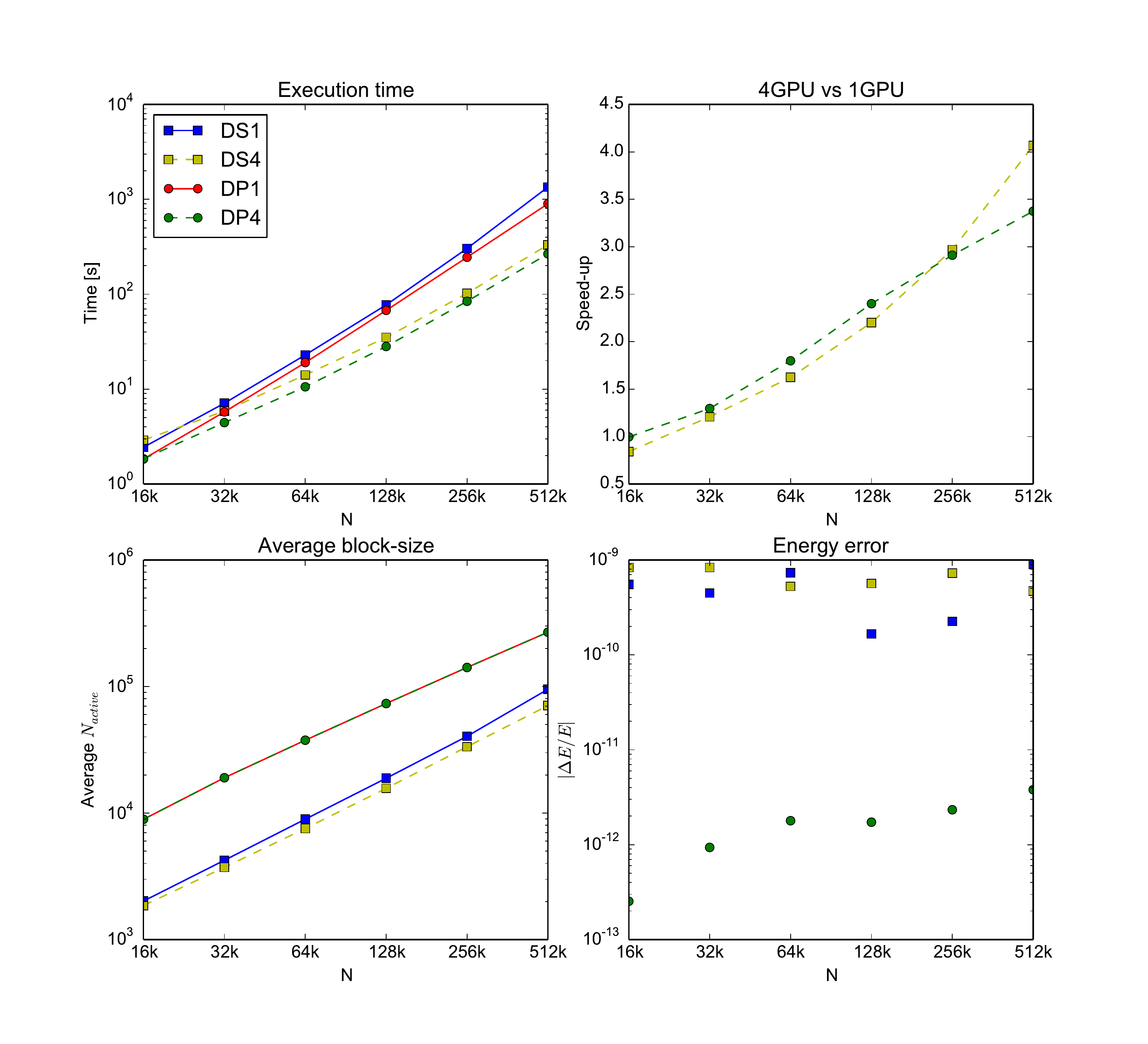}
\caption{Results of block time-step simulations with a sixth order hermite code.
The figure presents four sub-plots. For each of the plot the x-axis indicates 
the number of particles used. 
The four lines indicate the number of GPUs and the accuracy that is used. 
The solid blue line with square symbols uses double-single (DS) precision using 
a single GPU. The dashed yellow line with square symbols uses DS and four GPUs. 
The solid red line with round symbols uses full double precision (DP) using 
a single GPU. The dashed green line with round symbols uses DP and 
four GPUs.
The top left sub-plot presents the absolute execution time of the simulations.
The top right plot shows the speed-up when using four GPUs instead of one GPU.
The bottom left plot indicates the average number of particles that were 
being integrated per time-step ($N_\text{active}$).
The bottom right plot marks the energy error at the end of the simulation.
NOTE that the results for one and four GPUs when using DP are the same for 
the bottom two plots.
Timings performed on {\tt K20m} GPUs using {\tt CUDA 5.5}.}
\label{Sapporo2:fig:hermite6}
\end{figure}

\begin{figure}[h!]
  \centering  
  \includegraphics[width=\columnwidth,natwidth=360,natheight=360]{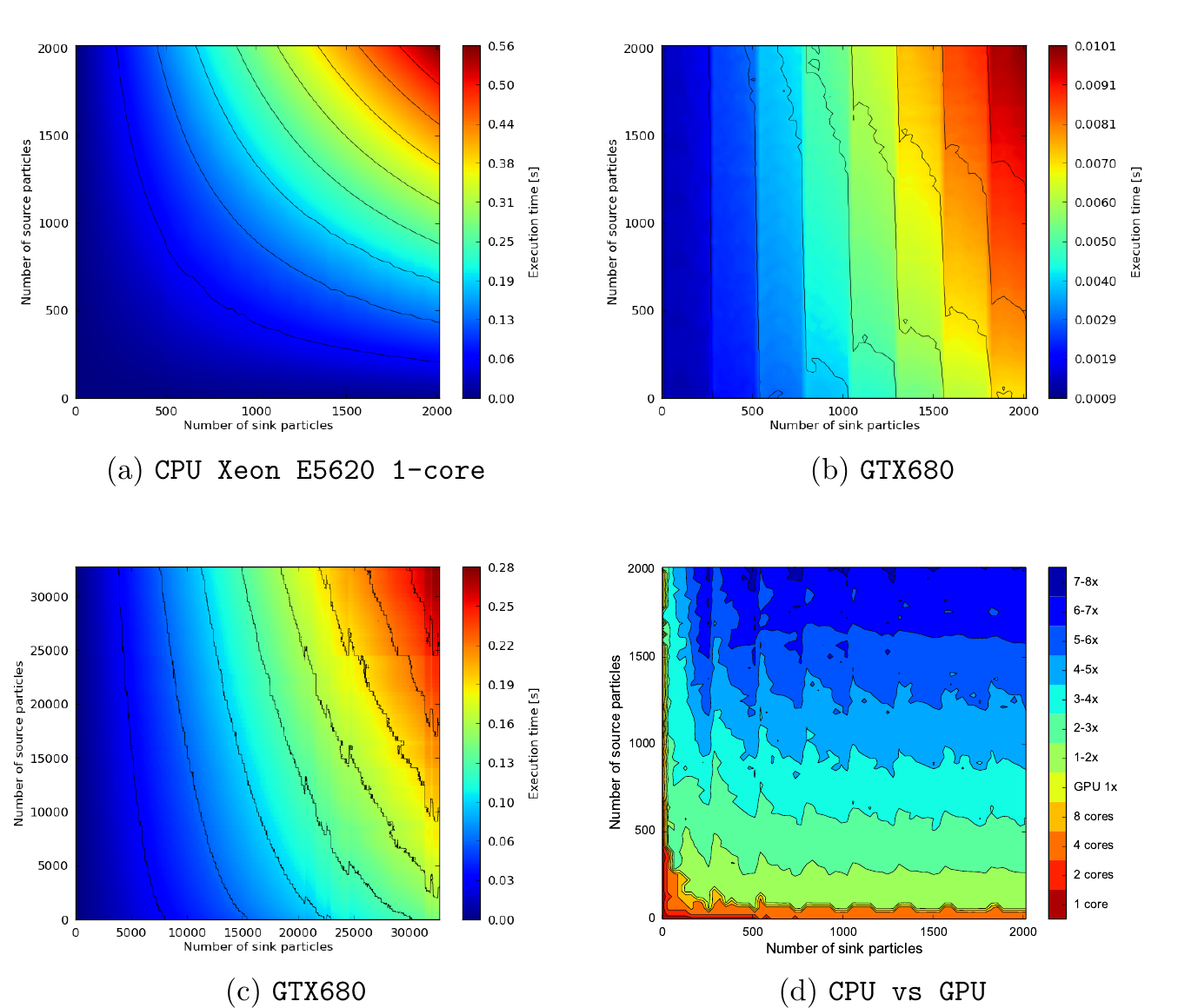}  
  \caption{GPU and CPU execution times. In all the subplots the x-axis indicates the number of sink
  particles and the y-axis the number of source particles used. For subplots a,b and c the raw 
  execution times are presented and indicated with the colours. Plot d does not present the 
  execution time but rather which of the configuration gives the best performance. If the GPU
  is faster than the 8 cores of the CPU we indicate by how much faster the GPU performs.
  To use more than a single CPU core we use OpenMP.
  Note that the colours are scaled per plot and are not comparable between the different 
  subplots. All the GPU times include the time required to copy data between the host and device. 
}
\label{Sapporo2:fig:CPUvsGPU}
\end{figure}

%
%


\section*{Tables}

\begin{table}[h!]
\renewcommand{\arraystretch}{1.3}
\small
\caption{GPUs used in this work. The first column indicates the GPU,
followed by three columns that show the memory properties. 
The clock-speed in Mhz in the second, the bus width in bits in the third
and the product of the two, the bandwidth in GB/s in the fourth.
The fifth column contains the number of compute cores and the sixth their 
clock-speed in Mhz. The next two columns indicate the theoretical
performance in TFlop/s, the single precision performance is in the seventh
column and the double precision in the eight column.
The next two columns gives the relative performance of each GPU where we set 
the GTX480 to 1. For the ninth column these numbers are determined using the 
theoretical peak single precision performance (TPP) of the chips.
The tenth column indicates the relative practical single precision peak performance (PPP) 
which is determined using a simple embarrassingly parallel $N$-body code.
}
\label{Sapporo2:Tab:GPUs}
\centering
\begin{tabular}{|l|ccc|cc|c|c|c|c|c|}    
\hline
     & \multicolumn{3}{c|}{Memory}  & \multicolumn{2}{c|}{Cores} & SP & DP  & TPP & PPP \\
     &  Mhz & bus & bw     & \# & Mhz & TFlop/s & TFlop/s  &  & \\
\hline
GTX480     & 3696 & 384 & 133.9  & 480   & 1401  & 1.35 & 0.17    &  1    & 1     \\\hline
GTX680     & 6008 & 256 & 192.2  & 1536  & 1006  & 3.09 & 0.13    &  2.3  & 1.7   \\\hline
K20m       & 5200 & 320 & 208    & 2496  & 706   & 3.5  & 1.17    &  2.6  & 1.8   \\\hline
GTX Titan  & 6144 & 384 & 288.4  & 2688  & 837   & 4.5  & 1.5     &  3.35 & 2.2   \\\hline
HD7970     & 5500 & 384 & 264    & 2048  & 925   & 3.8  & 0.94    &  2.8  & 2.3   \\\hline

\end{tabular}
\end{table}



%
%

\end{backmatter}
\end{document}